\documentclass[aps, pre, reprint, amsmath, amssymb, superscriptaddress, floatfix]{revtex4-2}

\usepackage{graphicx}
\usepackage{bm}
\usepackage{dcolumn}
\usepackage{xcolor}
\usepackage{booktabs}
\usepackage{placeins}

\begin{document}

\title{Statistical mechanics of the $N$-queens problem}

\author{Zong-Yue Liu}
\affiliation{Kuang Yaming Honors School, Nanjing University, Nanjing 210093, China}

\author{Hai-Jun Liao}
\affiliation{Institute of Physics, Chinese Academy of Sciences, Beijing 100190, China}

\author{Lei Wang}
\affiliation{Institute of Physics, Chinese Academy of Sciences, Beijing 100190, China}

\date{May 11, 2026}

\begin{abstract}
We investigate the $N$-queens problem as a lattice gas---a model in which $N$ queens
are placed on an $N \times N$ chessboard with pairwise repulsive
interactions along shared rows, columns, and diagonals---from the
perspective of statistical mechanics.  The
ground states are exactly the $Q(N)$ solutions of the classical
$N$-queens problem, with entropy per queen
$s_0 \approx \ln N - \gamma$ ($\gamma \approx 1.944$).
This entropy reflects a characteristic constraint hierarchy:
each successive geometric constraint---columns, then
diagonals---reduces the entropy from the free-placement value
$\ln N$ by a definite constant.
We derive the exact high-temperature energy $E/N \to 5/3$ as
$N \to \infty$.
Extensive Monte Carlo simulations with $10^8$ sweeps per temperature
point for $N = 8$--$1024$ reveal that the specific heat per queen
$C_v/N$ converges to a universal function of $T$ as $N \to \infty$.
The converged curve features a non-divergent peak
$C_v^{\max}/N \approx 1.63$ at $T^* \approx 0.235\,J$, establishing
the absence of a thermodynamic phase transition.
Combined with the trivially exact high-temperature entropy
$S(\infty)/N = \frac{1}{N}\ln\binom{N^2}{N}$, the
convergence of $C_v/N$ enables a thermodynamic integration of
$C_v/T$ from $T = \infty$ to $T = 0$ that recovers the ground-state
entropy---and hence the Simkin constant $\gamma$---purely from
Monte Carlo data.  This provides an independent
\emph{thermodynamic} route to a fundamental combinatorial constant.
Thermodynamic integration yields
$\gamma_{\rm MC} = 1.946 \pm 0.003$ at $N = 1024$, within $0.1\%$ of the
precise combinatorial value $\gamma = 1.94400(1)$.
We further present a transfer-matrix-based tensor network formulation
that encodes the non-attacking constraints into a rank-9 site tensor
with 17 nonzero elements, providing a complementary exact-enumeration
route.
\end{abstract}

\maketitle

\section{Introduction}
\label{sec:intro}

The $N$-queens problem---placing $N$ mutually non-attacking queens on
an $N \times N$ chessboard---is among the oldest combinatorial
problems in mathematics, dating back to the work of Bezzel in
1848~\cite{Bezzel1848}.  Despite its long history, the asymptotic
enumeration of solutions was resolved only recently.  Simkin~\cite{Simkin2022}
proved that the number of solutions $Q(N)$ satisfies
$Q(N) \sim (N/e^{\gamma})^N$.
This result builds on the breakthrough of Bowtell and
Keevash~\cite{Bowtell2023} and earlier bounds by Luria and
Simkin~\cite{Luria2021}.  Simkin's proof introduces \emph{queenons}---limit
objects defined as probability measures on the unit square---and
characterizes $\gamma$ as the solution of a convex optimization
problem over this space; the upper bound employs the entropy method
while the lower bound is established by a randomized construction
algorithm, yielding $\gamma \in [1.939,\, 1.945]$.
Nobel et al~\cite{Nobel2023} subsequently refined these
bounds to $\gamma = 1.94400(1)$ by solving
Simkin's convex programs with large-scale Newton methods.
On the computational side, exact enumeration of $Q(N)$ by exhaustive
backtracking search has been pushed to $N = 27$ using massively
parallel GPU computation~\cite{Yao2025}, highlighting the
combinatorial explosion that limits direct counting to moderate board
sizes.
All of these determinations are purely combinatorial.
In this work, we pursue an independent, thermodynamic route.
The starting point is the entropy per queen implied by the
asymptotic formula $Q(N) \sim (N/e^\gamma)^N$:
$s_0 = \lim_{N \to \infty} \ln Q(N)/N = \ln N - \gamma$.

The form of this entropy is suggestive.  On an
$N \times N$ board with one queen per row---the row constraint being
enforced by construction---each queen can occupy any of the $N$
columns, giving $\ln N$ units of entropy per queen.  The $N$-queens entropy $s_0 = \ln N - \gamma$ can
therefore be read as: each queen retains the full lattice entropy
$\ln N$, reduced by a constant $\gamma$ that encodes the cumulative
cost of geometric constraints.  This constant $\gamma$ decomposes
through a hierarchy of increasingly stringent geometric
constraints~\cite{Knuth2022}---column non-repetition costs one unit
of entropy (the Stirling correction), and the two diagonal constraints
cost a further $\gamma - 1 \approx 0.94$ units
(see Sec.~\ref{sec:entropy} for details).
That each constraint reduces the per-queen entropy by a well-defined
constant, independent of $N$ in the large-$N$ limit, hints at a
deeper statistical-mechanical structure underlying the combinatorics.

This additive decomposition of entropy into constraint costs motivates
a statistical-mechanical approach.  We map the $N$-queens
problem to a lattice gas with tunable
temperature~\cite{Zhang2009,Polson2024} and perform extensive Monte
Carlo simulations.  Throughout this paper we set $k_B = 1$, so that
entropy is dimensionless and temperature has units of energy.
A central finding is that the specific heat per
queen, $C_v/N$, converges to a size-independent curve for
sufficiently large $N$, exhibiting a Schottky-type peak but no phase
transition.  Combined with the exactly known high-temperature entropy
$S(\infty)/N = \frac{1}{N}\ln\binom{N^2}{N} \approx \ln N + 1$,
this convergence enables a thermodynamic integration
\begin{equation*}
  s_0 = \frac{S(\infty)}{N} - \int_0^\infty \frac{C_v}{NT}\,dT,
\end{equation*}
which extracts the ground-state entropy---and hence the Simkin
constant $\gamma$---purely from Monte Carlo simulation data.
This provides an independent \emph{thermodynamic} route to $\gamma$:
the only combinatorial input is the trivially exact high-temperature
entropy $S(\infty)/N = \frac{1}{N}\ln\binom{N^2}{N}$, while the
Simkin--Nobel value is used solely for post-hoc comparison.
Our thermodynamic extraction yields $\gamma_{\rm MC} = 1.946 \pm 0.003$ at
$N = 1024$ (within $0.1\%$ of the precise value~\cite{Nobel2023});
the monotonic convergence of $\gamma_{\rm MC}$ toward the asymptotic
value with increasing $N$ is shown in Fig.~\ref{fig:cv}(b).

We also present a complementary exact counting approach based on tensor
network contractions.  Each non-attacking constraint---row, column, or
diagonal---is encoded as a matrix product operator with bond
dimension $D = 2$.  Their intersection at every lattice site produces
a rank-9 local tensor with only 17 nonzero elements.  Exact
contraction of the resulting network yields $Q(N)$ without
statistical noise, providing independent benchmarks at moderate~$N$.

The paper is organized as follows.
Section~\ref{sec:model} defines the model.
Section~\ref{sec:analytical} presents analytical results: the
ground-state entropy and constraint hierarchy, and the exact
high-temperature limit.
Section~\ref{sec:mc} presents Monte Carlo results.
Section~\ref{sec:thermo_int} carries out the thermodynamic integration.
Section~\ref{sec:tensor} presents a tensor network formulation that
provides an alternative exact-enumeration route.
Section~\ref{sec:conclusions} concludes with a discussion of the limitations of the Monte Carlo and tensor network approaches and open problems.

\section{Model and simulation methodology}
\label{sec:model}

We consider an $N \times N$ square lattice with open boundary
conditions.  Each site $(i,j)$ ($i,j = 1,\ldots,N$) carries an
occupation variable $n_{ij} \in \{0,1\}$.  The Hamiltonian counts
mutually attacking queen pairs:
\begin{equation}
  H = J \sum_{\langle (i,j),(i',j') \rangle_{\rm attack}}
  n_{ij}\, n_{i'j'},
  \label{eq:H}
\end{equation}
where the sum runs over all pairs sharing a row ($i = i'$), column
($j = j'$), main diagonal ($i - j = i' - j'$), or anti-diagonal
($i + j = i' + j'$), and $J > 0$.  We set $J = 1$ throughout.

The lattice decomposes naturally into \textit{attack lines}: $N$ rows,
$N$ columns, $2N-1$ main diagonals, and $2N-1$ anti-diagonals.
Writing $n_\alpha$ for the number of queens on line $\alpha$, the
Hamiltonian takes the compact form
\begin{equation}
  H = \sum_{\alpha \in \mathcal{L}} \binom{n_\alpha}{2},
  \label{eq:H_lines}
\end{equation}
where $\mathcal{L}$ denotes the set of all $6N - 2$ attack lines and
$\binom{n}{2} = n(n-1)/2$ counts the number of attacking pairs on
each line.  This decomposition into line contributions is the
starting point for the high-temperature calculation
(Sec.~\ref{sec:highT}).

\begin{figure}[t]
  \includegraphics[width=0.75\linewidth]{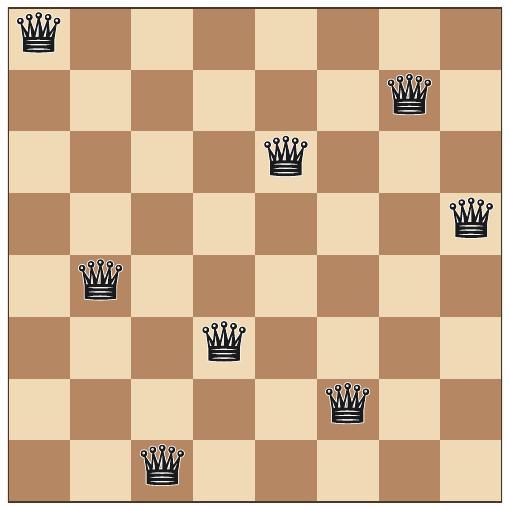}
  \caption{One of the $Q(8) = 92$ non-attacking ground-state
  configurations on an $8 \times 8$ chessboard.  No two queens share
  a row, column, or diagonal.}
  \label{fig:schematic}
\end{figure}

We work in the canonical ensemble with fixed particle number
$N$---the natural choice since each valid $N$-queens configuration
places exactly $N$ queens---and use Kawasaki dynamics~\cite{Kawasaki1966}
(queen--vacancy exchange) with the Metropolis acceptance
criterion~\cite{Metropolis1953}.  We choose queen--vacancy rather
than queen--queen exchanges because the former moves a single queen to
any empty site on the board, allowing large spatial displacements in
one step, whereas queen--queen swaps merely permute occupied sites and
cannot efficiently explore configurations with different spatial
structure.  One \textit{sweep} consists of $N$
attempted exchanges.  Statistical errors are estimated by jackknife
resampling with 200 bins~\cite{Efron1982}.
We simulate eight system sizes $N = 8, 16, 32, 64, 128, 256, 512, 1024$
on a uniform grid of 280 temperature points spanning $T = 0.05$ to
$500\,J$, with $10^8$ measurement sweeps and $2 \times 10^6$
thermalization sweeps at every temperature point.  The grid places
160 points in the specific-heat peak region ($T = 0.1025$--$0.500\,J$
in steps of $0.0025\,J$), 11 points in the low-temperature region
($T = 0.050$--$0.100\,J$, step $0.005\,J$), and the remaining 109
points at intermediate and high temperatures up to $T = 500\,J$.
All eight sizes use the same temperature grid and measurement
parameters, ensuring a self-consistent data set.
The simulations were carried out using 280 CPU cores in parallel,
with each core handling all eight system sizes at a single temperature
point in serial; the total wall-clock time was approximately
nine hours.
We verified ergodicity by running independent simulations from
multiple random initial configurations (seeds); all runs yielded
statistically consistent results for $E/N$ and $C_v/N$ at every
temperature, confirming that the Kawasaki dynamics explores the
configuration space adequately within our thermalization and
measurement windows.
A representative ground-state configuration is shown in
Fig.~\ref{fig:schematic}.

\section{Analytical results}
\label{sec:analytical}

\subsection{Ground-state entropy and constraint hierarchy}
\label{sec:entropy}

For $N$ queens, the ground states ($E = 0$) correspond to the
$Q(N)$ non-attacking configurations.  Since
$Q(N) \sim (N/e^\gamma)^N$, the entropy per queen is
\begin{equation}
  s_0 = \lim_{N\to\infty}\frac{\ln Q(N)}{N}
  = \ln N - \gamma, \quad \gamma = 1.94400(1).
  \label{eq:s0}
\end{equation}

The three-level constraint hierarchy (Table~\ref{tab:entropy_levels}),
implicit in the Stirling approximation and Simkin's
result~\cite{Simkin2022,Knuth2022},
provides a useful physical interpretation.  Starting from a fully
unconstrained model where each queen independently chooses one of $N$
columns ($\Omega_{\rm free} = N^N$, $s_{\rm free} = \ln N$), the
column non-repetition constraint reduces the count to $N!$
permutations.  By Stirling's approximation,
$\ln N! = N\ln N - N + O(\ln N)$, so the entropy per queen is
\begin{equation}
  s_{\rm perm} = \frac{\ln N!}{N}
  = \ln N - 1 + O\!\left(\frac{\ln N}{N}\right).
  \label{eq:s_perm}
\end{equation}
Thus the column constraint costs exactly one unit of entropy---the
Stirling correction arising from the non-repetition constraint on
column assignments.

The two diagonal constraints further reduce the entropy to
$s_0 = \ln N - \gamma$, costing an additional $\gamma - 1 \approx
0.94$ units, or approximately $(\gamma - 1)/2 \approx 0.47$ units
per diagonal family.  The hierarchy of costs is physically natural:
the column constraint, being a global permutation on $N$ equivalent
lines, removes exactly one unit of freedom per queen, while each
diagonal family---whose lines have unequal lengths ranging from $1$
to $N$---removes roughly half a unit.

\begin{table}[t]
  \caption{Entropy per queen under successive constraints for the
  $N$-queens problem.  The column constraint costs exactly 1 unit;
  the two diagonal constraints cost $\gamma - 1 \approx 0.94$ units
  combined.}
  \label{tab:entropy_levels}
  \begin{ruledtabular}
  \begin{tabular}{llcc}
    Model & Constraints & $\Omega$ & $s = \frac{1}{N}\ln\Omega$ \\
    \midrule
    Free & Row only & $N^N$ & $\ln N$ \\
    Permutation & Row + column & $N!$ & $\ln N - 1$ \\
    $N$-queens & Row + col + 2 diag & $\approx (N/e^\gamma)^N$
      & $\ln N - \gamma$ \\
  \end{tabular}
  \end{ruledtabular}
\end{table}

That each geometric constraint reduces the per-queen entropy by a
constant independent of $N$ suggests that the finite-temperature
thermodynamics may also admit a simple analytical description.  We
turn to this question next, beginning with the high-temperature limit.

\subsection{High-temperature limit}
\label{sec:highT}

At $T \to \infty$, all $\binom{N^2}{N}$ configurations are equally
probable.  Following Polson and Sokolov~\cite{Polson2024}, the exact
per-queen energy can be obtained by linearity of expectation over the
attack-line decomposition [Eq.~\eqref{eq:H_lines}].  The probability
that any two given sites are simultaneously occupied is
$P = N(N{-}1)/[N^2(N^2{-}1)]$, and the total number of attacking
site pairs summed over all $6N{-}2$ lines is
$S_{\rm tot} = N(N{-}1)(5N{-}1)/3$.  Combining these gives the exact
result
\begin{equation}
  \frac{\langle E \rangle}{N}\bigg|_{T \to \infty}
  = \frac{(5N-1)(N-1)}{3N(N+1)}
  \xrightarrow{N \to \infty} \frac{5}{3},
  \label{eq:E_highT}
\end{equation}
valid for all finite $N$.  Our Monte Carlo simulations
(Fig.~\ref{fig:energy}) confirm this: at high $T$, $E/N$ converges to
Eq.~\eqref{eq:E_highT} with corrections of order $O(1/N)$.

\section{Finite-temperature Monte Carlo results}
\label{sec:mc}

\subsection{Convergence diagnostics}

\begin{figure}[t]
  \includegraphics[width=\linewidth]{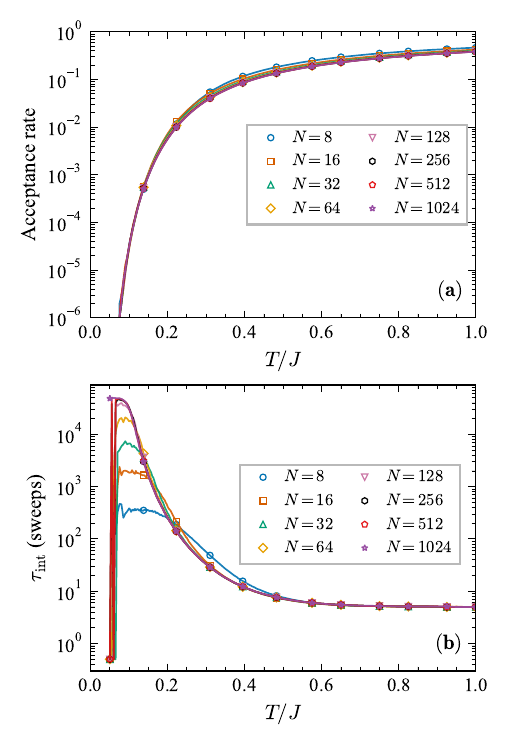}
  \caption{Convergence diagnostics for the Monte Carlo simulations.
  (a)~Acceptance rate of the Kawasaki (queen--vacancy exchange) moves
  as a function of temperature for $N = 8$--$1024$ (log scale).
  The rate drops below $10^{-5}$ for $T \lesssim 0.1\,J$ and rises
  monotonically, reaching $0.38$--$0.47$ at $T = 1\,J$.
  (b)~Integrated autocorrelation time $\tau_{\rm int}$ of the energy
  in units of sweeps (log scale).  The sharp peak at
  $T \approx 0.055$--$0.10\,J$ reflects low-temperature freezing and
  lies well below the specific heat maximum $T^* \approx 0.235\,J$;
  at $T^*$ itself, $\tau_{\rm int} \approx 85$--$200$ sweeps,
  yielding more than $5 \times 10^5$ independent samples from $10^8$
  sweeps.}
  \label{fig:convergence}
\end{figure}

Before presenting thermodynamic observables, we verify that the
Monte Carlo simulations have reached equilibrium and that the
measurements are statistically independent.
Figure~\ref{fig:convergence} shows the acceptance rate and the
integrated autocorrelation time $\tau_{\rm int}$ of the energy as
functions of temperature for several system sizes.

The acceptance rate [panel~(a)] increases monotonically with $T$,
reaching $\sim 0.38$--$0.47$ at $T = 1\,J$; it drops below $10^{-5}$
for $T \lesssim 0.1\,J$, reflecting the near-perfect freezing into
non-attacking ground states.  For $N \ge 32$ the curves collapse,
confirming convergence to the thermodynamic limit.

The autocorrelation time [panel~(b)] exhibits a sharp peak at
$T \approx 0.055$--$0.10\,J$, well below the specific heat maximum
$T^* \approx 0.235\,J$, reaching $\tau_{\rm int} \approx 5.0 \times
10^4$ sweeps for $N = 1024$.  Even at this worst case, $10^8$
measurement sweeps yield $\sim 2\,000$ independent samples---adequate
for the essentially featureless low-temperature regime where both
$E/N$ and $C_v/N$ approach zero exponentially.  Crucially,
at the physically important specific heat peak ($T \approx 0.235\,J$),
$\tau_{\rm int} \approx 85$--$200$ sweeps for all $N$, providing
more than $5 \times 10^5$ independent samples per temperature point
and ensuring high statistical precision in the region that matters
most.  At $T/J > 0.4$, $\tau_{\rm int}$ falls to $\sim 5$--$10$
sweeps for all system sizes.
We additionally verified ergodicity by comparing independent runs
from different random initial configurations, all yielding
statistically consistent results.

\subsection{Energy}

\begin{figure}[t]
  \includegraphics[width=\linewidth]{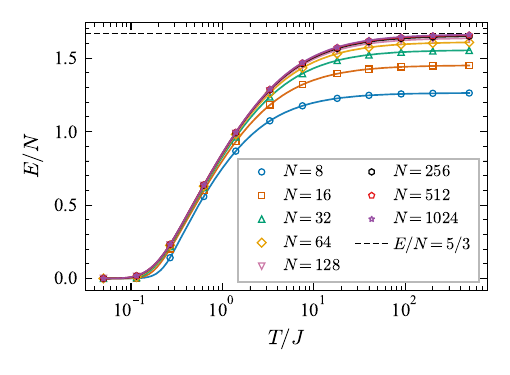}
  \caption{Energy per queen $E/N$ vs.\ temperature (logarithmic
  scale) for $N = 8$--$1024$ over the full range
  $T = 0.05$--$400\,J$.  The dashed line marks the analytical
  $N \to \infty$ high-$T$ limit $E/N = 5/3$.}
  \label{fig:energy}
\end{figure}

Figure~\ref{fig:energy} shows $E/N$ vs.\ $T$ for $N = 8$--$1024$.
The curves for $N \ge 32$ are nearly indistinguishable, confirming
rapid convergence to the thermodynamic limit.  At high temperatures,
$E/N$ approaches the exact finite-$N$ value
[Eq.~\eqref{eq:E_highT}], which itself converges to $5/3$ as
$N \to \infty$.

\subsection{Specific heat convergence}

\begin{figure}[t]
  \includegraphics[width=\linewidth]{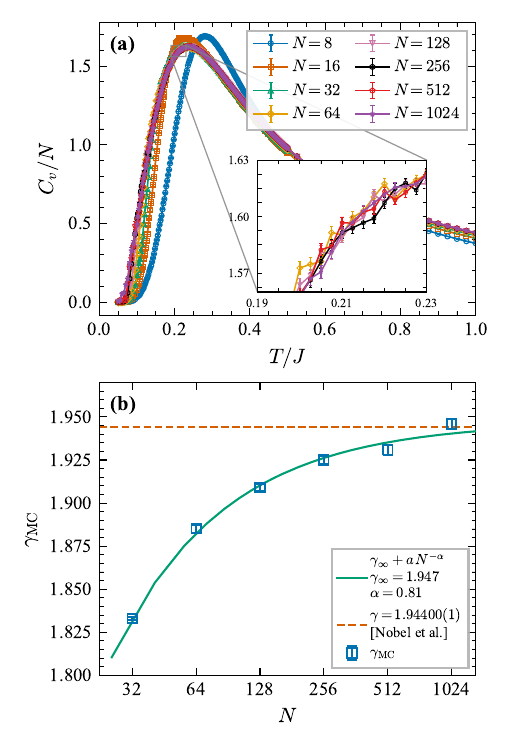}
  \caption{(a)~Specific heat per queen $C_v/N$ vs.\ $T$ for
  $N = 8$--$1024$ in the peak region ($T \le 1\,J$, linear scale)
  with error bars.  Inset: zoom into the peak for $N \ge 64$.
  The curves for $N \ge 32$ collapse onto a single function,
  confirming convergence to the thermodynamic limit.
  (b)~Extracted Simkin constant
  $\gamma_{\rm MC} = \ln N - s_0^{\rm MC}$ vs.\ $N$ for
  $N = 32$--$1024$.  The dashed line marks the precise combinatorial
  value $\gamma = 1.94400(1)$~\cite{Nobel2023}.  The solid curve is
  an ad hoc finite-size scaling fit
  $\gamma_{\rm MC}(N) = \gamma_\infty + a\,N^{-\alpha}$, yielding
  $\gamma_\infty = 1.947 \pm 0.004$ and $\alpha = 0.81 \pm 0.06$.
  Jackknife-propagated error bars on the data points are plotted but
  are smaller than or comparable to the symbol size.}
  \label{fig:cv}
\end{figure}

The specific heat per queen $C_v/N = (\langle E^2\rangle - \langle
E\rangle^2)/(T^2 N)$ is shown in Fig.~\ref{fig:cv}(a).  For $N \ge 32$,
the $C_v/N$ curves collapse onto a single function of $T$, indicating
that the thermodynamic limit has been reached.
Table~\ref{tab:cv_peak} lists the peak values.

\begin{table}[t]
  \caption{$C_v/N$ peak scaling.
  For $N = 8$--$128$, the peak values are from dedicated simulations
  with dense temperature sampling in $T \in (0.20,\,0.30)\,J$;
  for $N = 256$--$1024$, values are from the standard 280-point grid.
  The peak height saturates near $1.63$ for $N \ge 32$, consistent with
  the absence of a thermodynamic phase transition.}
  \label{tab:cv_peak}
  \begin{ruledtabular}
  \begin{tabular}{rD{.}{.}{4}D{.}{.}{4}D{.}{.}{4}}
    $N$ & \multicolumn{1}{c}{$T_{\rm peak}$}
    & \multicolumn{1}{c}{$C_v^{\max}/N$}
    & \multicolumn{1}{c}{error} \\
    \midrule
    8   & 0.281 & 1.6915 & 0.0002 \\
    16  & 0.227 & 1.6783 & 0.0006 \\
    32  & 0.231 & 1.6288 & 0.0005 \\
    64  & 0.237 & 1.6216 & 0.0004 \\
    100 & 0.237 & 1.6229 & 0.0004 \\
    128 & 0.237 & 1.6225 & 0.0004 \\
    \midrule
    256  & 0.2450 & 1.6238 & 0.0020 \\
    512  & 0.2400 & 1.6287 & 0.0023 \\
    1024 & 0.2325 & 1.6251 & 0.0024 \\
  \end{tabular}
  \end{ruledtabular}
\end{table}

The key observation is that $C_v^{\max}/N$ saturates near $1.63$ for
$N \ge 32$ and shows no systematic growth with system size
(Table~\ref{tab:cv_peak}).  The variation across
$N = 32$--$1024$ is $\Delta C_v^{\max}/N \lesssim 0.007$, far smaller
than the $\sim 0.06$ per doubling expected from a logarithmic
divergence $\propto \ln N$.
This behavior is consistent with the absence of a thermodynamic phase
transition and identifies the crossover as a
Schottky-type anomaly~\cite{Binder2010}---a broad peak in $C_v$
arising from thermal excitation across a finite energy gap, without
any symmetry breaking.  Physically, the peak reflects thermal
excitation from the non-attacking ground states ($E = 0$) to
configurations with a small number of attacking pairs
($E \sim O(1)$); the finite energy gap between these two regimes
produces the characteristic broad maximum without critical
fluctuations.  The peak temperature $T^* \approx 0.235\,J$
stabilizes for $N \ge 64$.

The convergence of $C_v/N$ to a definite limiting function has a
powerful consequence: since both the high-temperature entropy
$S(\infty)/N$ and the ground-state entropy $s_0$ are independently
known, the integral $\int_0^\infty C_v/(NT)\,dT$ must equal their
difference.  We exploit this in the next section.

\section{Thermodynamic integration}
\label{sec:thermo_int}

The convergence of $C_v/N$ to a universal function enables a powerful
consistency check.  We compute the entropy
difference between $T = \infty$ and $T = 0$ in two independent ways
and verify that they agree.

\textit{Combinatorial route.}---At $T \to \infty$, all configurations
are equally probable, so the entropy per queen is
\begin{equation}
  \frac{S(\infty)}{N}
  = \frac{1}{N}\ln\binom{N^2}{N}.
  \label{eq:Sinf_def}
\end{equation}
For $N \gg 1$, applying the Stirling approximation
$\ln n! = n\ln n - n + \tfrac{1}{2}\ln(2\pi n) + O(n^{-1})$:
\begin{align}
  \ln\binom{N^2}{N}
  &= \ln(N^2!) - \ln N! - \ln(N^2 - N)! \notag\\
  &= N\ln N - (N^2{-}N)\ln(1{-}1/N)
  + \Delta_{\rm S},
  \label{eq:Sinf_stirling}
\end{align}
where the first two terms come from the $n\ln n - n$ part of Stirling's
formula and the correction
$\Delta_{\rm S} = \tfrac{1}{2}\ln[N^2/(2\pi N(N^2-N))]
= -\tfrac{1}{2}\ln(2\pi(N-1))$ collects the subleading logarithms.
Expanding $\ln(1 - 1/N) = -1/N - 1/(2N^2) - \cdots$, the
leading terms give $-(N^2-N)\ln(1-1/N) = N - \tfrac{1}{2} + O(N^{-1})$,
so
\begin{equation}
  \frac{S(\infty)}{N}
  = \ln N + 1 + O\!\left(\frac{\ln N}{N}\right).
  \label{eq:Sinf}
\end{equation}
At $T = 0$, only ground states contribute:
$S(0)/N = \ln Q(N)/N \approx \ln N - \gamma$.

\textit{Thermodynamic route.}---The fundamental relation
\begin{equation}
  S(\infty) - S(0) = \int_0^\infty \frac{C_v}{T}\,dT
  \label{eq:thermo_int}
\end{equation}
relates the entropy difference to the specific heat.  Dividing by $N$
and substituting the results above:
\begin{equation}
  \int_0^\infty \frac{C_v}{NT}\,dT
  = \frac{S(\infty) - S(0)}{N}
  = 1 + \gamma + O\!\left(\frac{\ln N}{N}\right).
  \label{eq:DS}
\end{equation}
This relation connects a purely thermodynamic quantity---the
integrated specific heat on the left---with purely combinatorial
constants on the right: the Stirling correction (contributing unity)
and the Simkin constant $\gamma$.

Rearranging Eq.~\eqref{eq:DS}, we can extract the ground-state
entropy per queen directly from the Monte Carlo data:
\begin{equation}
  s_0^{\rm MC}
  = \frac{S(\infty)}{N} - \int_0^\infty \frac{C_v}{NT}\,dT,
  \label{eq:s0_MC}
\end{equation}
where $S(\infty)/N = \frac{1}{N}\ln\binom{N^2}{N}$ is the trivially
exact entropy of the uniform distribution over all $\binom{N^2}{N}$
configurations---requiring no knowledge of the ground-state count
$Q(N)$ or the Simkin constant.
This provides a purely thermodynamic determination of the
ground-state degeneracy without direct enumeration; the Simkin--Nobel
value $\gamma = 1.94400(1)$ enters only as a benchmark for comparison.

The numerical integration uses 280 temperature grid points
spanning from $T = 0.05\,J$ to $400\,J$; the low-$T$
cutoff contributes negligibly since $C_v/T$ is exponentially
suppressed below $T \approx 0.1\,J$, and the high-$T$ cutoff at
$400\,J$ is sufficient because $C_v/T \propto T^{-3}$ decays
rapidly, contributing less than $0.01\%$ of the integral beyond
this point.  The trapezoidal rule on the logarithmic grid introduces
discretization errors well below the statistical uncertainty of the
Monte Carlo data.

Table~\ref{tab:entropy_check} summarizes the results.
For $N = 8$ and $N = 16$, where exact values of $Q(N)$ are known
($Q(8) = 92$, $Q(16) = 14\,772\,512$), $s_0^{\rm MC}$ agrees with
the exact $s_0^{\rm exact} = \ln Q(N)/N$ to within $0.1\%$,
validating the simulation and integration procedure.

For $N \ge 32$, where exact $Q(N)$ is unknown, we use $s_0^{\rm MC}$
to extract the Simkin constant via
$\gamma_{\rm MC} = \ln N - s_0^{\rm MC}$.
The extracted values converge toward the known
asymptotic value $\gamma = 1.94400(1)$~\cite{Nobel2023}:
$\gamma_{\rm MC} = 1.833 \pm 0.001$ ($N = 32$), $1.885 \pm 0.002$ ($N = 64$),
$1.909 \pm 0.002$ ($N = 128$), $1.925 \pm 0.002$ ($N = 256$),
$1.931 \pm 0.003$ ($N = 512$), and $1.946 \pm 0.003$ ($N = 1024$), with the deviation
decreasing from $5.7\%$ to $0.11\%$ at $N = 1024$.
The dominant source of error at moderate $N$ is the finite-size
correction to the Simkin formula, which makes $\gamma_{\rm eff}(N)
= \ln N - s_0(N)$ systematically smaller than the asymptotic
$\gamma$; the MC integration itself introduces only sub-percent
errors, as confirmed by the $N = 8$ and $N = 16$ benchmarks.
The statistical uncertainties on $s_0^{\rm MC}$ and $\gamma_{\rm MC}$
(Table~\ref{tab:entropy_check}) are obtained by propagating jackknife
errors on the individual $C_v(T)$ data points through the trapezoidal
integration.  At $N = 1024$, the propagated uncertainty
$\sigma(\gamma_{\rm MC}) = 0.003$ is small compared to the $0.11\%$
deviation from the asymptotic value, confirming that the dominant
source of error is the finite-size correction to the Simkin formula
rather than statistical noise.

A finite-size scaling extrapolation
(e.g., $\gamma_{\rm MC}(N) = \gamma_\infty + a\,N^{-\alpha}$) could in
principle improve the infinite-$N$ estimate.  Fitting the six data
points $N = 32$--$1024$ to this three-parameter form yields
$\gamma_\infty = 1.947 \pm 0.004$ with $\alpha \approx 0.81$
[Fig.~\ref{fig:cv}(b), solid curve], consistent with the precise
value $\gamma = 1.94400(1)$.  However, the functional form of the
subleading corrections to $Q(N)$ is not known analytically, making
this fitting form ad hoc; the extrapolated value should therefore
be regarded as indicative rather than definitive.

The monotonic convergence of $\gamma_{\rm MC}$ toward $\gamma = 1.94400(1)$ with
increasing $N$---reaching $0.11\%$ accuracy at $N = 1024$---supports
the mutual consistency of the Monte Carlo
thermodynamics, the Stirling approximation for $S(\infty)$, and the
Simkin combinatorial formula.

\begin{table}[t]
  \caption{Ground-state entropy from thermodynamic integration.
  $s_0^{\rm MC} = S(\infty)/N - \int_0^\infty (C_v/NT)\,dT$, where
  $S(\infty)/N = \frac{1}{N}\ln\binom{N^2}{N}$ is exact.
  Statistical uncertainties are obtained by propagating jackknife
  errors on $C_v(T)$ through the trapezoidal integration.
  For $N \le 16$, $s_0^{\rm MC}$ is compared with the exact
  ground-state entropy $s_0^{\rm exact} = \ln Q(N)/N$.
  For $N \ge 32$, we extract
  $\gamma_{\rm MC} = \ln N - s_0^{\rm MC}$ and compare with the
  precise value $\gamma = 1.94400(1)$~\cite{Nobel2023}.
  All data use a 280-point temperature grid with $10^8$
  measurement sweeps per point.}
  \label{tab:entropy_check}
  \begin{ruledtabular}
  \begin{tabular}{ccccc}
    $N$ & $s_0^{\rm MC}$
    & $s_0^{\rm exact}$
    & $\gamma_{\rm MC}$
    & Deviation \\
    \midrule
    8   & $0.566 \pm 0.000$ & 0.565 & --- & 0.1\% \\
    16  & $1.031 \pm 0.001$ & 1.032 & --- & 0.1\% \\
    \midrule
    32  & $1.633 \pm 0.001$ & --- & $1.833 \pm 0.001$ & 5.7\% \\
    64  & $2.274 \pm 0.002$ & --- & $1.885 \pm 0.002$ & 3.0\% \\
    128 & $2.943 \pm 0.002$ & --- & $1.909 \pm 0.002$ & 1.8\% \\
    256 & $3.620 \pm 0.002$ & --- & $1.925 \pm 0.002$ & 1.0\% \\
    512 & $4.308 \pm 0.003$ & --- & $1.931 \pm 0.003$ & 0.69\% \\
    1024 & $4.985 \pm 0.003$ & --- & $1.946 \pm 0.003$ & 0.11\% \\
  \end{tabular}
  \end{ruledtabular}
\end{table}

\section{Tensor network formulation}
\label{sec:tensor}
The geometric constraints of the $N$-queens problem---specifically,
the requirement that no two queens share the same row, column, or diagonal---can be
precisely encoded using the matrix product operator (MPO) formalism~\cite{Frowis2010}.
Consider, for instance, a single row with 3 sites. Without imposing any constraints,
the row admits $2^3$ possible configurations, corresponding to
$(|0\rangle + |1\rangle)^{\otimes 3}$, where $|0\rangle$ and
$|1\rangle$ denote the absence and presence of a queen at each site, respectively.
To enforce the condition that each row contains exactly one queen, we construct an MPO as follows:
\begin{align}
  & \hat{n}_1 \otimes \hat{n}_0 \otimes \hat{n}_0 
  + \hat{n}_0 \otimes \hat{n}_1 \otimes \hat{n}_0 
  + \hat{n}_0 \otimes \hat{n}_0 \otimes \hat{n}_1 \nonumber \\
  & =  \bm{v}_0 * A * A * A * \bm{v}_1^T = M_{\rm row},
\end{align}
where the operator $\hat{n}_0 = |0\rangle\langle 0|$ projects onto
the empty state (no queen), and $\hat{n}_1 = |1\rangle\langle 1|$
projects onto the occupied state (one queen). The local tensor $A$ of the MPO is
\begin{equation}
  A = \begin{pmatrix}
    \hat{n}_0 & \hat{n}_1 \\
    0         & \hat{n}_0  
  \end{pmatrix}, 
\end{equation}
and the boundary vectors $\bm{v}_0 = (1, 0)$ and $\bm{v}_1 = (0, 1)$
indicate that each row starts with zero queens and ends with exactly
one queen, respectively. Applying this MPO to the unconstrained state
space directly produces a ground state wavefunction that contains all
configurations in which each row contains exactly one queen, i.e.,
\begin{equation} 
|\Psi \rangle =  (\bm{v}_0 * A * A * A * \bm{v}_1^T) (|0\rangle + |1\rangle)^{\otimes 3}.
\end{equation}
Similarly, the constraint that there is \emph{at most one} queen along
any (anti-)diagonal can also be encoded as an MPO
\begin{align}
  &\,\,\,\,\,\,\,\, \hat{n}_1 \otimes \hat{n}_0 \otimes \hat{n}_0 
   + \hat{n}_0 \otimes \hat{n}_1 \otimes \hat{n}_0 
   + \hat{n}_0 \otimes \hat{n}_0 \otimes \hat{n}_1 \nonumber \\
 & + \hat{n}_0 \otimes \hat{n}_0 \otimes \hat{n}_0 =  \bm{v}_0 * A * A * A * \bm{v}_2^T = M_{\rm diag},
\end{align}
The only modification compared to the ``exactly one'' case is
that the boundary vector at the end, previously chosen as
$\bm{v}_1 = (0, 1)$, should instead be $\bm{v}_2 = (1, 1)$. This
choice allows for zero or one queen on the constraint line, thus
correctly implementing the ``at most one'' requirement for diagonals
and anti-diagonals within the MPO framework.

\begin{figure}[t]
  \includegraphics[width=\linewidth]{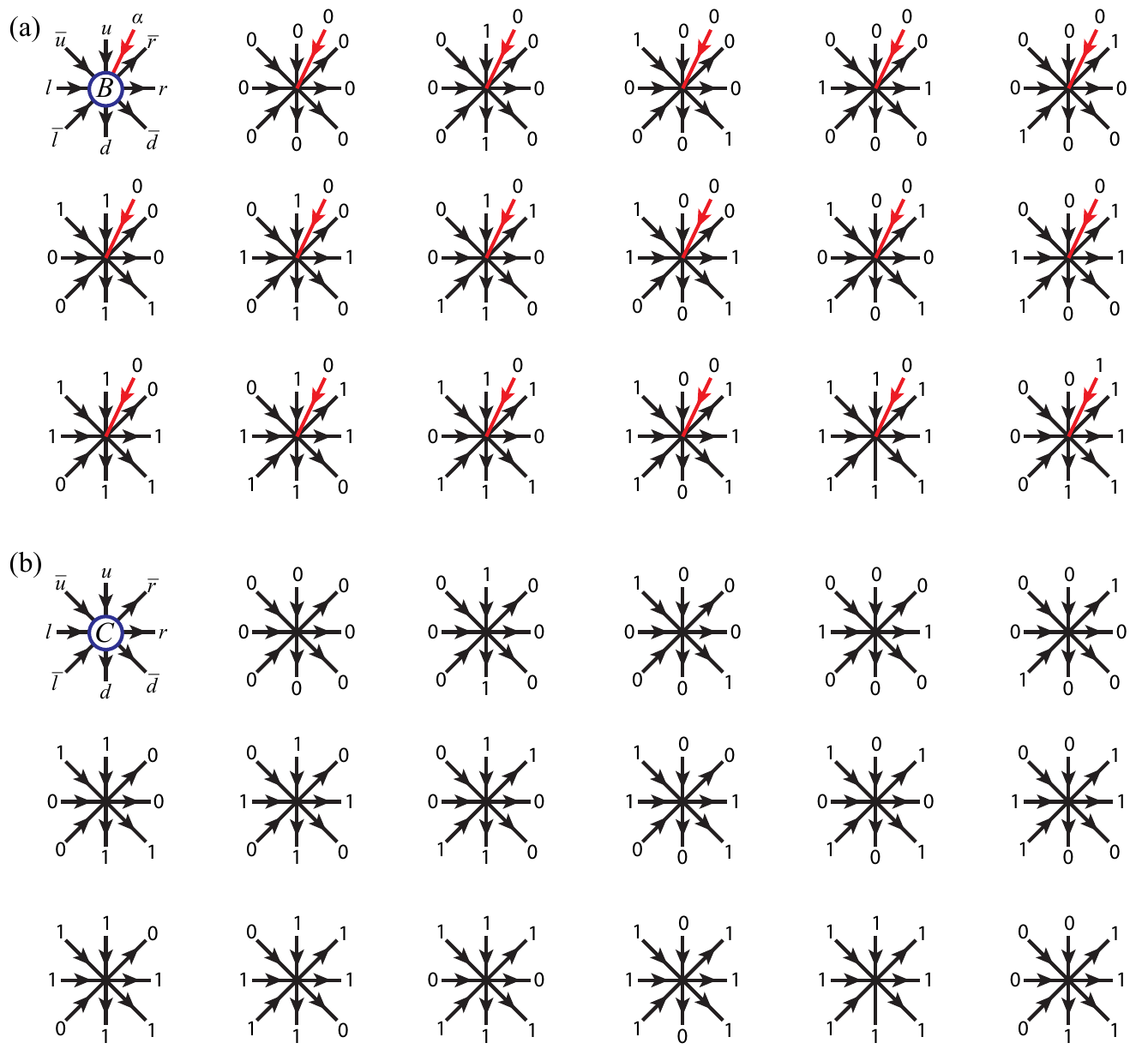}
  \caption{(a)~The 17 nonzero elements of the rank-9 site tensor $B$
  [Eq.~\eqref{eq:Btensor}].
  Each diagram shows a lattice site with eight bond indices
  carrying binary attack signals and one physical index
  $\alpha \in \{0,1\}$ (red arrows); the first 16 elements
  correspond to empty sites ($\alpha = 0$) with all $2^4$
  combinations of independent pass-through signals along the four
  constraint channels (column, row, and two diagonal families),
  and the 17th element (bottom right) represents an occupied site
  ($\alpha = 1$): all incoming signals are $0$ and all outgoing
  signals are $1$.
  (b)~The 17 nonzero elements of the rank-8 site tensor $C$
  [Eq.~\eqref{eq:Ctensor}], obtained by summing $B$ over the
  physical index $\alpha$.  This is the local tensor used in the
  tensor network contraction for computing $Q(N)$.}
  \label{fig:tensor}
\end{figure}

By combining all MPOs that enforce constraints for rows, columns,
and diagonals, we obtain the ground state wavefunction
that contains every solution configuration of the $N$-queens problem.
Since each site $(i,j)$ lies at the intersection of four constraint lines,
the ground state wavefunction $|\Psi \rangle$ can be written as a tensor network state with
a rank-9 local tensor $B$ at each site, namely,
\begin{align}
 B_{ud,\,lr,\,\bar{u}\bar{d},\,\bar{l}\bar{r}}^{\,\alpha}
  & = \sum_{\alpha'\beta'\beta\sigma} A_{ud}^{\alpha\alpha'}\;A_{lr}^{\alpha'\beta'}\;A_{\bar{u}\bar{d}}^{\beta'\beta}\;
    A_{\bar{l}\bar{r}}^{\beta\sigma} |\sigma \rangle, \nonumber \\
  & = A_{ud}^{\alpha\alpha}\;A_{lr}^{\alpha\alpha}\;A_{\bar{u}\bar{d}}^{\alpha\alpha}\;A_{\bar{l}\bar{r}}^{\alpha\alpha}\,|\alpha\rangle,
  \label{eq:Btensor}
\end{align}
where $(u,d)$, $(l,r)$, $(\bar{u},\bar{d})$, $(\bar{l},\bar{r})$ are the virtual indices that
carry binary signals along the column, row, $\searrow$-diagonal, and
$\nearrow$-diagonal directions, $\alpha \in \{0,1\}$ is the physical index labeling the site occupation,
and $\alpha',\beta',\beta,\sigma \in\{0,1\}$ are internal summation indices
that collapse to $\alpha$ because $A$ is diagonal in the physical space.  Since the local tensor $A_{ij}^{\sigma\sigma'}$ of the MPO has only three nonzero elements, namely,
$A_{00}^{00} = A_{11}^{00} = 1$ (empty site: signal pass-through)
and $A_{01}^{11} = 1$ (occupied site: emit attack signal), the local tensor $B$ of the ground state
wavefunction $|\Psi \rangle$ has exactly 17 nonzero elements (see Fig.~\ref{fig:tensor}).
For an empty site ($\alpha = 0$), each of the four channels independently
passes or blocks its signal, giving $2^4 = 16$ elements.
For an occupied site ($\alpha = 1$), all incoming signals must vanish and
all outgoing signals are emitted, giving one additional element.

\begin{figure}[t]
  \includegraphics[width=\linewidth]{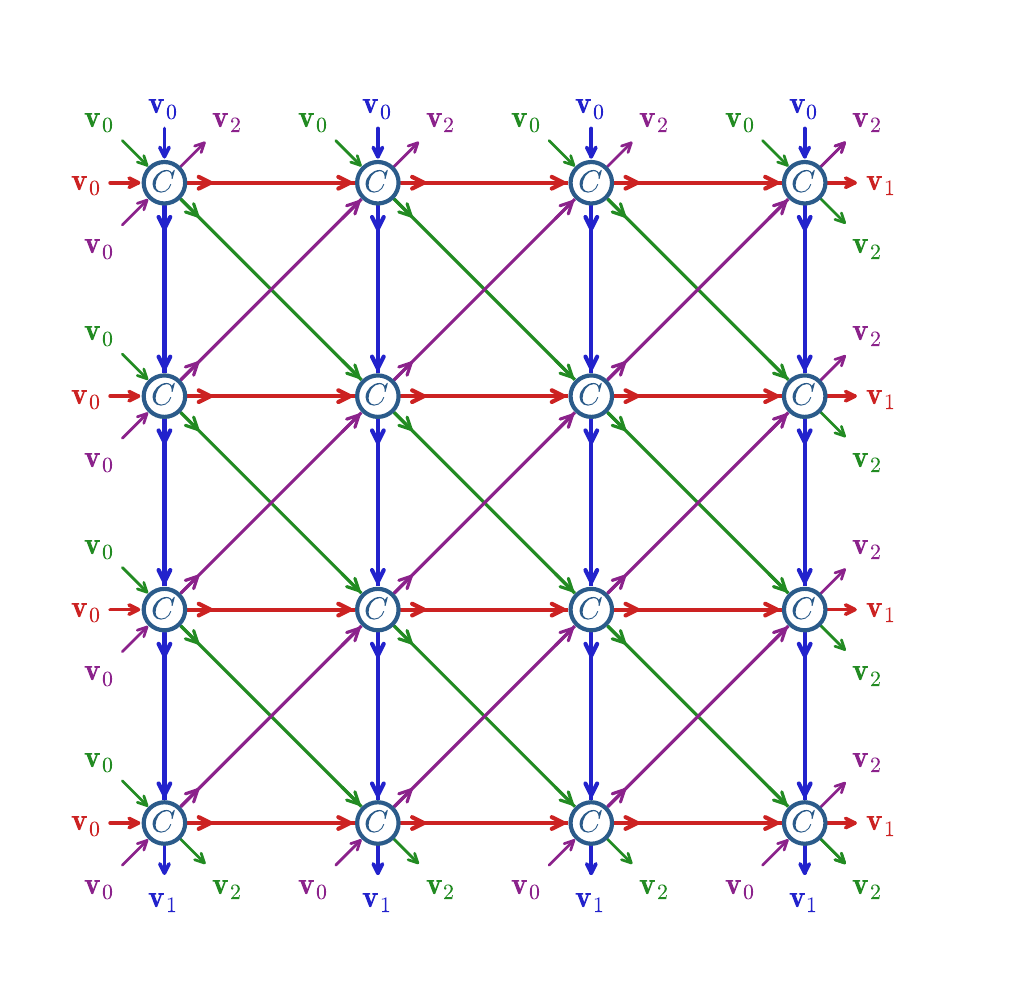}
  \caption{Tensor network for exact counting of $N$-queens solutions
  ($N = 4$).  Each node is the rank-8 site tensor~$C$
  [Eq.~\eqref{eq:Ctensor}].
  Arrows on each bond indicate the signal propagation direction:
  red (row), blue (column), green ($\searrow$-diagonal),
  purple ($\nearrow$-diagonal).
  Boundary vectors $\mathbf{v}_0 = (1,0)$, $\mathbf{v}_1 = (0,1)$,
  and $\mathbf{v}_2 = (1,1)$ are labeled at every line terminus,
  enforcing the queen constraints on each row, column, and diagonal.}
  \label{fig:tn_network}
\end{figure}

Since the ground state wavefunction $|\Psi \rangle$ contains all solution 
configurations of the $N$-queens problem, the exact number of solutions $Q(N)$ 
is obtained by 
\begin{equation}
Q(N) = \langle \Psi | (|0 \rangle +  |1 \rangle)^{\otimes N^2},
\end{equation}
where $(|0 \rangle +  |1 \rangle)^{\otimes N^2}$ enumerates all possible 
occupation configurations. Since $(|0 \rangle +  |1 \rangle)^{\otimes N^2}$ 
is a product state, it is equivalent to contracting each physical index with 
the vector $(1,1)^T = |0\rangle + |1\rangle$. As a result, the exact number
of solutions $Q(N)$ can be obtained by contracting a tensor network with a 
bond dimension $D=2$ local tensor $C$ at each site, namely,
\begin{equation}
  C_{ud,\,lr,\,\bar{u}\bar{d},\,\bar{l}\bar{r}} = \sum_{\alpha} B_{ud,\,lr,\,\bar{u}\bar{d},\,\bar{l}\bar{r}}^{\,\alpha},
   \label{eq:Ctensor}
  \end{equation}
 which also has 17 nonzero elements (as shown in Fig.~\ref{fig:tensor}~(b)).
Figure~\ref{fig:tn_network} illustrates the complete tensor network for 
the 4-queens problem.

More generally, tensor network contraction provides a powerful
framework for exact counting in constraint satisfaction
problems~\cite{Kourtis2019,Vanderstraeten2018}, of which the $N$-queens problem is a
natural instance.
The tensor network can be exactly contracted row by row using
boundary matrix product state methods~\cite{Xiang2024}.  The bond
dimension of the boundary MPS grows exponentially with~$N$, so the
computational cost is exponential, but the results are free of
statistical noise.  This makes exact contraction a natural source of
benchmarks that complement the Monte Carlo data at moderate~$N$.

Another perspective on the tensor network contraction is the
row-by-row transfer matrix $\mathcal{T}$.  Contracting the network
row by row with appropriate boundary vectors $|v_0\rangle$ and
$\langle v_f|$ gives
\begin{equation}
  Q(N) = \langle v_f |\,\mathcal{T}^N\,| v_0 \rangle,
\end{equation}
where $|v_0\rangle = \mathbf{v}_0^{\otimes N} \otimes
\mathbf{v}_0^{\otimes (2N-1)} \otimes \mathbf{v}_0^{\otimes (2N-1)}$
is the product of initial boundary vectors for all columns and
diagonals (each initialized to $\mathbf{v}_0 = (1,0)$, signaling no
queen seen), and $\langle v_f|$ projects onto the state in which every
column carries exactly one queen ($\mathbf{v}_1 = (0,1)$) while each
diagonal carries at most one ($\mathbf{v}_2 = (1,1)$).
A crucial property of $\mathcal{T}$ is that it is
\emph{nilpotent}: $\mathcal{T}^{N+1} = 0$.
This follows from the column constraint.  Each application of
$\mathcal{T}$ processes one row and can only increase the number of
occupied columns; since there are at most $N$ columns, after $N+1$
applications all states are annihilated.  Consequently, every
eigenvalue of $\mathcal{T}$ is exactly zero.

This nilpotent structure stands in sharp contrast to conventional
statistical-mechanical models.  For the two-dimensional Ising model,
for instance, the transfer matrix has a unique largest eigenvalue
$\lambda_1 \sim \alpha^N$ for some constant $\alpha$, and the
partition function $Z \sim \lambda_1^N \sim \alpha^{N^2}$ is
controlled by this eigenvalue.  Standard infinite-size contraction
techniques such as CTMRG~\cite{Nishino1996} exploit precisely this
spectral dominance.  For the $N$-queens transfer matrix, however,
there is no nonzero eigenvalue to target: the solution count
$Q(N)$ resides entirely in the matrix elements of $\mathcal{T}^N$,
not in its spectrum.  This rules out spectral methods (DMRG, VUMPS,
CTMRG) and makes finite-$N$ exact contraction the natural framework,
as described above.


\FloatBarrier
\section{Conclusions}
\label{sec:conclusions}

We have demonstrated that the Simkin constant $\gamma$---a
fundamental quantity in combinatorics---can be determined through
an independent thermodynamic route.  By formulating the $N$-queens
problem as a lattice gas and performing extensive Monte Carlo
simulations ($N = 8$--$1024$, $10^8$ sweeps per temperature point),
we map out the specific heat $C_v/N$ over the entire temperature
range.  A key prerequisite is that $C_v/N$ converges to a
size-independent function for $N \ge 32$, with a non-divergent
Schottky-type peak at $T^* \approx 0.235\,J$ consistent with the
absence of a thermodynamic phase transition.

This convergence, combined with the trivially exact high-temperature
entropy $S(\infty)/N = \frac{1}{N}\ln\binom{N^2}{N}$, enables a
thermodynamic integration of $C_v/T$ that yields the Simkin constant
as $\gamma_{\rm MC} = \ln N - S(\infty)/N + \int_0^\infty
(C_v/NT)\,dT$.  The method is validated at $N = 8$ and $16$ where
exact solution counts are known (agreement within $0.1\%$), and for
larger $N$ produces $\gamma_{\rm MC} = 1.946 \pm 0.003$ at
$N = 1024$, within $0.11\%$ of the precise combinatorial
value $\gamma = 1.94400(1)$~\cite{Nobel2023}.  The remaining deviation
is dominated by finite-size corrections to the Simkin formula rather
than statistical noise, and extending to still larger system sizes
would further sharpen the estimate.

These results establish the $N$-queens lattice gas as an interesting 
problem from the statistical mechanics perspective. Monte Carlo
simulation can be used to estimate the
Simkin constant $\gamma$.
The tensor network formulation presented in Sec.~\ref{sec:tensor}
provides a complementary exact-enumeration route: it yields $Q(N)$
without statistical noise at moderate~$N$.  Together, the two methods
cover complementary regimes of~$N$. An open direction is to
exploit whether the tools in computational quantum and statistical mechanics will 
offer new results in this ancient problem in combinatorics.

\section{Data availability}
The simulation data and analysis code are publicly available on
GitHub~\cite{github_repo}.

\section{Acknowledgments}
The authors acknowledge valuable discussions with Jinguo Liu, Yijia Wang, Qi Yang, Pan Zhang, Youjin Deng and Tao Xiang. This work is supported by the National Natural Science Foundation of China under Grants No. T2225018, No. 12188101, No. T2121001, the Cross-Disciplinary Key Project of Beijing Natural Science Foundation No. Z250005, the Strategic Priority Research Program of the Chinese Academy of Sciences under Grants No. XDB0500000, and the National Key Projects for Research and Development of China Grants No. 2021YFA1400400.


\end{document}